\documentclass{emulateapj}
\usepackage{apjfonts}
\usepackage{rotating}

\newcommand{\pivec}{\mbox{\boldmath $\pi$}}
\newcommand{\muvec}{\mbox{\boldmath $\mu$}}

\lefthead{TSAPRAS ET AL.} 
\righthead{A REFINED ANALYSIS OF MICROLENSING EVENT OGLE-2012-BLG-0406}

\begin{document}

\title{A SUPER-JUPITER ORBITING A LATE-TYPE STAR: A REFINED ANALYSIS OF MICROLENSING EVENT OGLE-2012-BLG-0406.}

\author{
Y. Tsapras$^{1,R1,\spadesuit}$,
J.-Y. Choi$^{K1}$,
R. A. Street$^{1,\spadesuit}$,
C. Han$^{K1,{\star},\ddag}$,
V. Bozza$^{N3,\diamondsuit}$,
A. Gould$^{U1,\ddag}$,
M. Dominik\footnote{Royal Society University Research Fellow}$^{R3,\diamondsuit,\spadesuit,\clubsuit}$,
J.-P. Beaulieu$^{P4,\clubsuit}$,
A. Udalski$^{O1,\blacksquare}$,
U. G. J{\o}rgensen$^{N1,\diamondsuit}$
% ...
\\(The MOA Collaboration)
}

\affil{$^{1}$Las Cumbres Observatory Global Telescope Network, 6740 Cortona Drive, suite 102, Goleta, CA 93117, USA}
\affil{$^{K1}$Department of Physics, Chungbuk National University, Cheongju 361-763, Republic of Korea}

\affil{$^{R1}$School of Physics and Astronomy, Queen Mary University of London, Mile End Road, London E1 4NS, UK}
% ...
\affil{$^{R3}$SUPA, School of Physics \& Astronomy, University of St Andrews, North Haugh, St Andrews KY16 9SS, UK}
% ...
\affil{$^{N1}$Niels Bohr Institute, Astronomical Observatory, Juliane Maries vej 30, 2100 Copenhagen, Denmark}
% ...
\affil{$^{N3}$Dipartimento di Fisica ``E. R. Caianiello'', Universit\`a di Salerno,
Via Giovanni Paolo II n.~132, 84084 Fisciano (SA), Italy}
% ...
\affil{$^{O1}$ Warsaw University Observatory, Al. Ujazdowskie 4, 00-478 Warszawa, Poland}
% ...
\affil{$^{P4}$UPMC-CNRS, UMR7095, Institut d'Astrophysique de Paris, 98bis boulevard Arago, 75014 Paris, France}
% ...
\affil{$^{U1}$Department of Astronomy, Ohio State University, 140 West 18th Avenue, Columbus, OH 43210, USA}
% ...
\affil{$^{\spadesuit}$The RoboNet Collaboration}
\affil{$^{\diamondsuit}$The MiNDSTEp Collaboration}
\affil{$^{\blacksquare}$The OGLE Collaboration}
\affil{$^{\ddag}$The $\mu$FUN Collaboration}
\affil{$^{\clubsuit}$The PLANET Collaboration}
\affil{$^{\dag}$The MOA Collaboration\\}
\affil{$^{\star}$Corresponding author}

\begin{abstract}
We present a detailed analysis of survey and follow-up observations of microlensing 
event OGLE-2012-BLG-0406 based on data obtained from 10 different observatories. 
Intensive coverage of the lightcurve, especially the perturbation part, allowed us 
to accurately measure the parallax effect and lens orbital motion.  Combining our 
measurement of the lens parallax with the angular Einstein radius determined from 
finite-source effects, we estimate the physical parameters of the lens system. 
We find that the event was caused by a $2.73\pm 0.43\ M_{\rm J}$ planet orbiting 
a $0.44\pm 0.07\ M_{\odot}$ early M-type star.  The distance to the lens is 
$4.97\pm 0.29$\ kpc and the projected separation between the host star and its 
planet at the time of the event is $3.45\pm 0.26$ AU. We find that the additional coverage 
provided by follow-up observations, especially during the planetary perturbation, 
leads to a more accurate determination of the physical parameters of the lens.
\end{abstract}

\keywords{gravitational lensing -- binaries: general -- planetary systems}

\section{Introduction}
Radial velocity and transit surveys, which primarily target main-sequence 
stars, have already discovered hundreds of giant planets and are now beginning 
to explore the reservoir of  lower mass planets with orbit sizes extending to 
a few astronomical units (AU). These planets mostly lie well inside the snow 
line\footnote{The snow line is defined as the distance from the star in a protoplanetary disk where
ice grains can form \citep{2006ApJ...640.1115L}.} of their host stars. Meanwhile, direct imaging with large 
aperture telescopes has been discovering giant planets tens to hundreds of AUs 
away from their stars \citep{2005Natur.435.1067K}. The region of sensitivity of microlensing lies somewhere 
in between and extends to low-mass exoplanets lying beyond the snow-line of their 
low-mass host stars, between $\sim$1 and 10 AU \citep{2003MNRAS.343.1131T, 2012ARA&A..50..411G}. 
Although there is already strong evidence  that cold sub-Jovian planets are more 
common than originally thought around low-mass stars
\citep{2006ApJ...644L..37G, 2010ApJ...710...1641, 2013A&A...552A..70K, 2013ApJS..204...24B}, cold 
super-Jupiters orbiting K or M-dwarfs were believed to be a rarer class of 
objects\footnote{although a metal-rich protoplanetary disk might allow the 
formation of sufficiently massive solid cores.} \citep{2004ApJ...612L..73L, 2011MNRAS.417..314M, 2012Natur.481..167C}.

Both gravitational instability and core accretion models of planetary formation have a hard time generating these planets, although it is possible to produce them given appropriate initial conditions. The main argument against core accretion is that it takes too long to produce a massive planet but this crucially depends on the core mass and the opacity of the planet envelope during gas accretion. In the case of gravitational instability, a massive protoplanetary disc would probably have too high an opacity to fragment locally at distances of a few AU.

The radial velocity method has been remarkably successful in tabulating the part of 
the distribution that lies within the snow-line but discoveries of super-Jupiters 
beyond the snow-line of M-dwarfs have been comparatively few \citep{2010PASP..122..149J, TRENDS-IV}. 
Since microlensing is most sensitive to planets that are further away from their host stars, typically M and K dwarfs, the two techniques are complementary \citep{2012ARA&A..50..411G}.

Three brown dwarf and nineteen planet microlensing discoveries have been published 
to date, including the discoveries of two multiple-planet systems 
\citep{2008Sci...319..927G, 2013ApJ...762L..28H}\footnote{For a complete list 
consult http://exoplanet.eu/catalog/ and references therein.}. It is also worth 
noting that unbound objects of planetary mass have also been reported \citep{2011Natur.473..349S}.

Microlensing involves the chance alignment along an observer's line of sight of 
a foreground object (lens) and a background star (source). This results in a 
characteristic variation of the brightness of the background source as it is 
being gravitationally lensed. As seen from the Earth, the brightness of the 
source increases as it approaches the lens, reaching a maximum value at the 
time of closest approach. The brightness then decreases again as the source 
moves away from the lens.

In microlensing events, planets orbiting the lens star can reveal their presence 
through distortions in the otherwise smoothly varying standard single lens lightcurve. 
Together, the host star and planet constitute a binary lens. Binary lenses have a 
magnification pattern that is more complex than the single lens case due to the presence 
of extended caustics that represent the positions on the source plane at which the 
lensing magnification diverges. Distortions in the lightcurve arise when the trajectory 
of the source star approaches (or crosses) the caustics \citep{1991ApJ...374L..37M}. Recent reviews of the method can be found in \cite{2010GReGr..42.2075D} and \cite{2011exop.book...79G}.

Upgrades to the OGLE\footnote{http://ogle.astrouw.edu.pl} \citep{2003AcA...53...291U} survey observing setup
and MOA\footnote{http://www.phys.canterbury.ac.nz/moa} \citep{2003ApJ...591..204S} microlensing 
survey telescope in the past couple of years brought greater precision and enhanced observing cadence, resulting in an 
increased rate of exoplanet discoveries. For example, OGLE has regularly been monitoring the field of the OGLE-2012-BLG-0406 event since March 2010 with a cadence of 55 minutes. When a microlensing alert was issued notifying the 
astronomical community that event OGLE-2012-BLG-0406 was exhibiting anomalous behavior, 
intense follow-up observations from multiple observatories around the world were initiated 
in order to better characterize the deviation. This event was first analyzed by
\cite{POLESKI} using exclusively the OGLE-IV survey photometry. That study concluded that the event was caused by a planetary system consisting of a 3.9$\pm$1.2 $M_{\rm J}$ planet orbiting a low mass late K/early M dwarf.
 
In this paper we present the analysis of the event based on the combined data obtained from 
10 different telescopes, spread out in longitude, providing dense and continuous coverage 
of the lightcurve. 

The paper is structured as follows: Details of the discovery of this event, follow-up 
observations and image analysis procedures are described in Section 2. Section 3 presents 
the methodology of modeling the features of the lightcurve. We provide a summary and 
conclude in Section 4.


\begin{deluxetable*}{lllr}
\tablecaption{Observations\label{table:one}}
\tablewidth{0pt}
\tablehead{
\multicolumn{1}{c}{group}     &
\multicolumn{1}{c}{telescope} & 
\multicolumn{1}{c}{passband} &
\multicolumn{1}{c}{data points} 
}
\startdata
OGLE          &  1.3m Warsaw Telescope, Las Campanas Observatory (LCO), Chile                           &  $I$      & 3013 \\
RoboNet       &  2.0m Faulkes North Telescope (FTN), Haleakala, Hawaii, USA                             &  $I$      & 83 \\
RoboNet       &  2.0m Faulkes South Telescope (FTS), Siding Spring Observatory (SSO), Australia         &  $I$      & 121 \\
RoboNet       &  2.0m Liverpool Telescope (LT), La Palma, Spain                                         &  $I$      & 131 \\
MiNDSTEp      & 1.5m Danish Telescope, La Silla, Chile                                                  &  $I$      & 473\\
MOA           &  0.6m Boller \& Chivens (B\&C), Mt. John, New Zealand                                   &  $I$      & 1856\\
$\mu$FUN      &  1.3m SMARTS, Cerro Tololo Inter-American Observatory (CTIO), Chile                     &  $V$, $I$ & 16, 81\\
PLANET        &  1.0m Elizabeth Telescope, South African Astronomical Observatory (SAAO), South Africa  &  $I$      & 226 \\
PLANET        &  1.0m Canopus Telescope, Mt. Canopus Observatory, Tasmania, Australia                   &  $I$      & 210 \\
WISE          & 1.0m Wise Telescope, Wise Observatory, Israel                                           &  $I$      & 180
\enddata  
\end{deluxetable*}

\section{Observations and data}
Microlensing event OGLE-2012-BLG-0406 was discovered at equatorial coordinates 
$\alpha=17^{\rm h}53^{\rm m}18.17^{\rm s}$, $\delta=-30\arcdeg28\arcmin16.2\arcsec$ 
(J2000.0)\footnote{$(l, b)=-0.46^{\circ},-2.22^{\circ}$} by the OGLE-IV survey and announced 
by their Early Warning System (EWS)\footnote{http://ogle.astrouw.edu.pl/ogle4/ews/ews.html} 
on the 6$^{\rm th}$ of April 2012. The event had a baseline $I$-band magnitude of 16.35 and 
was gradually increasing in brightness. The predicted maximum magnification at the time of 
announcement was low, therefore the event was considered a low-priority target for most 
follow-up teams who preferentially observe high-magnification events as they are associated 
with a higher probability of detecting planets \citep{1998ApJ...500...37G}. 

OGLE observations of the event were carried out with the
1.3-m Warsaw telescope at the Las Campanas Observatory, Chile, equipped
with the 32 chip mosaic camera. The event's field was visited every 55
minutes providing very dense and precise coverage of the entire light
curve from the baseline, back to the baseline. For more details on the
OGLE data and coverage see \cite{POLESKI}.

An assessment of data acquired by the OGLE team until the 1$^{\rm st}$ of July (08:47 UT, 
HJD$\sim$2456109.87) which was carried out by the SIGNALMEN anomaly detector \citep{SIGNALMEN} 
on the 2$^{\rm nd}$ of July (02:19 UT) concluded that a microlensing anomaly, i.e.\ a deviation 
from the standard bell-shaped Paczy\'{n}ski curve \citep{1986ApJ...304....1P}, was in progress. 
This was electronically communicated via the ARTEMiS (Automated Robotic Terrestrial Exoplanet
Microlensing Search) system \citep{2008AN....329..248D} to trigger prompt observations by both 
the RoboNet-II\footnote{http://robonet.lcogt.net} collaboration \citep{2009AN...330...4T} and the 
MiNDSTEp\footnote{http://www.mindstep-science.org} consortium \citep{2010AN....331..671D}. 
RoboNet's web-PLOP system \citep{PLOP} reacted to the trigger by scheduling observations already 
from the 2$^{\rm nd}$ of July (02:30 UT), just 11 minutes after the SIGNALMEN assessment started. 
However, the first RoboNet observations did not occur before the 4$^{\rm th}$ of July (15:26 UT), 
when the event was observed with the FTS. This delayed response was due to the telescopes being 
offline for engineering work and bad weather at the observing sites. It fell to the Danish 1.54m 
at ESO La Silla to provide the first data point following the anomaly alert (2$^{\rm nd}$ of July, 
03:42 UT) as part of the MiNDSTEp efforts. The alert also triggered automated anomaly modeling 
by RTModel \citep{2010MNRAS.408.2188B}, which by the 2$^{\rm nd}$ of July (04:22 UT) delivered 
a rather broad variety of solutions in the stellar binary or planetary range, reflecting the 
fact that the true nature was not well-constrained by the data available at that time. This 
process chain did not involve any human interaction at all. 

The first human involvement was an e-mail circulated to all microlensing teams by V.~Bozza 
on the 2$^{\rm nd}$ of July (07:26 UT) informing the community about the ongoing anomaly 
and modeling results. Including OGLE data from a subsequent night, the apparent anomaly 
was also independently spotted by E.~Bachelet (e-mail by D.P.~Bennett of 3$^{\rm rd}$ July, 
13:42 UT), and subsequently PLANET\footnote{http://planet.iap.fr} team \citep{2006Natur.439..437B} 
SAAO data as well as  $\mu$FUN\footnote{http://www.astronomy.ohio-state.edu/$\sim$microfun} 
\citep{2006ApJ...644L..37G} SMARTS (CTIO) data were acquired the coming night, which along 
with the RoboNet FTS data cover the main peak of the anomaly. It should be noted that the 
observers at CTIO decided to follow the event even while the moon was full in order to obtain 
crucial data. A model circulated by T.~Sumi on the 5$^{\rm th}$ of July (00:38 UT) did not 
distinguish between the various solutions. 

However, when the rapidly changing features of the anomaly were independently assessed by 
the Chungbuk National University group (CBNU, C. Han),  the community was informed on the 
5$^{\rm th}$ of July (10:43 UT) that the anomaly is very likely due to the presence of a 
planetary companion. An independent modeling run by V.~Bozza's automatic software (5$^{\rm th}$ 
of July, 10:55 UT) confirmed the result. While the OGLE collaboration (A. Udalski) notified 
observers on the 5$^{\rm th}$ of July that a caustic exit was occurring, a geometry leading 
to a further small peak successively emerged from the models. D.P.~Bennett circulated a model 
using updated data on the 6$^{\rm th}$ of July (00:14 UT) which highlighted the presence of 
a second prominent feature expected to occur $\sim$10$^{\rm th}$ of July. Another modeling 
run performed at CBNU on the 7$^{\rm th}$ of July (02:39 UT) also identified this feature 
and estimated that the secondary peak would occur on the 11$^{\rm th}$ of July.

Follow-up teams continued to monitor the progress of the event intensively until the beginning 
of September, well after the planetary deviation had ceased, and provided dense coverage of 
the main peak of the event. A preliminary model using available OGLE and follow-up data at 
the time, circulated on the 31$^{\rm st}$ of October (C. Han, J.-Y. Choi), classified the 
companion to the lens as a super-Jupiter. \cite{POLESKI} presented an analysis of this event 
using reprocessed survey data exclusively. In this paper we present a refined analysis using 
survey and follow-up data together.

The groups that contributed to the observations of this event, along with the telescopes used, 
are listed in Table~\ref{table:one}. Most observations were obtained in the $I$-band and some 
images were also taken in other bands in order to create a color-magnitude diagram and classify 
the source star. We note that there are also observations obtained from the MOA 1.8m survey 
telescope which we did not include in our modeling because the target was very close to the 
edge of the CCD. We also do not include data from the $\mu$FUN Auckland 0.4m, PEST 0.3m, 
Possum 0.36m and Turitea 0.36m telescopes due to poor observing conditions at site.

Extracting accurate photometry from observations of crowded fields, such as the Galactic Bulge, 
is a challenging process. Each image contains thousands of stars whose stellar point-spread 
functions (PSFs) often overlap so aperture and PSF-fitting photometry can at best offer limited 
precision. In order to optimize the photometry it is necessary to use difference imaging (DI) 
techniques \citep{1998ApJ...503..325A}. For any particular telescope/camera combination, DI 
uses a reference image of the event taken under optimal seeing conditions which is then degraded 
to match the seeing conditions of every other image of the event taken from that telescope. The 
degraded reference image is then subtracted from the matching image to produce a residual (or 
difference) image. Stars that have not varied in brightness in the time interval between the 
two images will cancel, leaving no systematic residuals on the difference image but variable 
stars will leave either a positive or negative residual. 

DI is the preferred method of photometric 
analysis among microlensing groups and each group has developed custom pipelines to reduce their 
observations. OGLE and MOA images were reduced using the pipelines described in \cite{2003AcA...53...291U} 
and \cite{2001MNRAS.327..868B} respectively.  PLANET, $\mu$FUN, and WISE images were processed 
using variants of the PySIS \citep{2009MNRAS.397.2099A} pipeline, whereas RoboNet and MiNDSTEp 
observations were analyzed using customized versions of the DanDIA package \citep{2008MNRAS...386...L77}. 
Once the source star returned to its baseline magnitude, each data set was reprocessed to optimize 
photometric precision. These photometrically optimized data sets were used as input for our 
modeling run.

\begin{figure*}[ht]
\epsscale{0.9}
\plotone{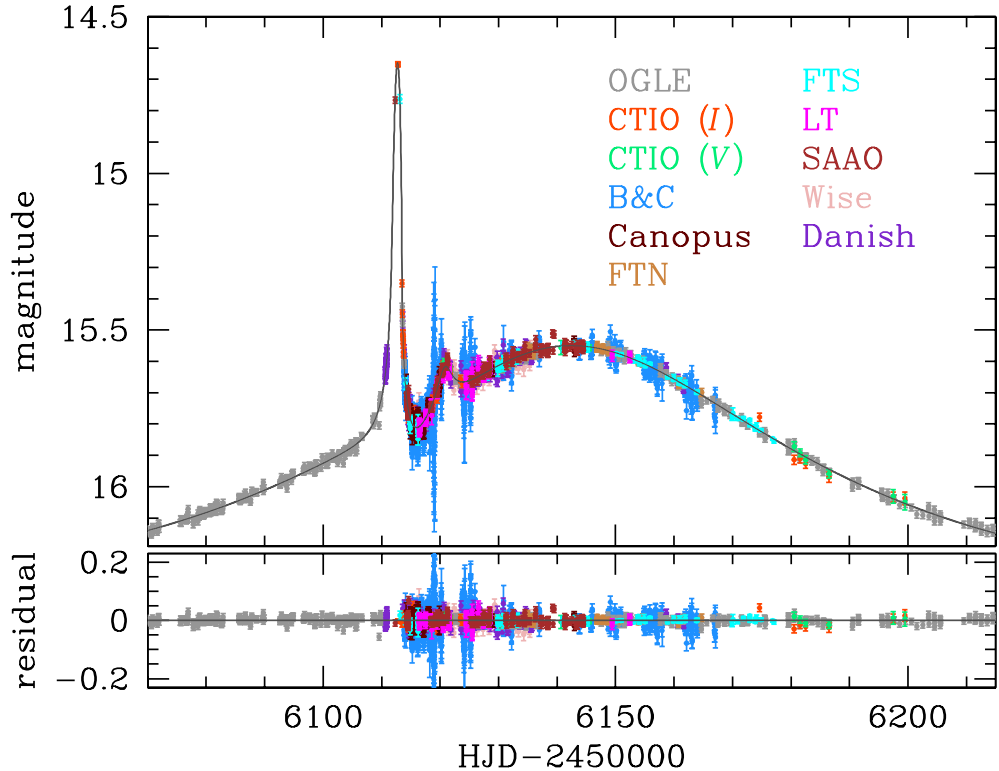}
\caption{\label{fig:one}
Lightcurve of OGLE-2012-BLG-0406 showing our best-fit binary-lens model including parallax and orbital motion.
The legend on the right of the figure lists the contributing telescopes. All data were taken in the $I$-band, 
except where otherwise indicated.
}
\end{figure*}


\section{Modeling}
Figure~\ref{fig:one} shows the lightcurve of OGLE-2012-BLG-406. The lightcurve displays two main features 
that deviate significantly from the standard Paczy\'{n}ski curve. The first feature, which peaked at 
HJD $\sim$ 2456112 (3$^{\rm rd}$ of July), is produced by the source trajectory grazing the cusp of a caustic. 
The brightness then quickly drops as the source moves away from the cusp \citep{1992A&A...260....1S, 1995A&A...293....1Z}, increases again for a brief 
period as it passes close to another cusp at HJD $\sim$ 2456121 (12$^{\rm th}$ of July), and eventually returns 
to the standard shape as the source moves further away from the caustic structure. The anomalous behavior, 
when both features are considered, lasts for a total of $\sim$ 15 days, while the full duration of the event 
is $\gtrsim$120 days. These are typical lightcurve features expected from lensing phenomena involving planetary lenses.

\begin{deluxetable*}{lrrrrrrr}
\tablecaption{Lensing Parameters\label{table:two}}
\tablewidth{0pt}
\tablehead{
\multicolumn{1}{l}{parameters} &
\multicolumn{1}{c}{standard} &
\multicolumn{2}{c}{parallax} &
\multicolumn{2}{c}{orbit} &
\multicolumn{2}{c}{orbit+parallax} \\
\multicolumn{1}{c}{} &
\multicolumn{1}{c}{} &
\multicolumn{1}{c}{$u_0>0$} &
\multicolumn{1}{c}{$u_0<0$} &
\multicolumn{1}{c}{$u_0>0$} &
\multicolumn{1}{c}{$u_0<0$} &
\multicolumn{1}{c}{$u_0>0$} &
\multicolumn{1}{c}{$u_0<0$}
}
\startdata
$\chi^2$/dof              & 6921.019/6383       & 6850.358/6381       & 6677.685/6381       & 6408.371/6381       & 6408.255/6381       & 6357.680/6379       & 6381.358/6379       \\
$t_0$ (HJD')              & 6141.63 $\pm$ 0.04  & 6141.70 $\pm$ 0.05  & 6141.66 $\pm$ 0.05  & 6141.24 $\pm$ 0.05  & 6141.28 $\pm$ 0.04  & 6141.33 $\pm$ 0.05  & 6141.19 $\pm$ 0.06  \\
$u_0$                     & 0.532 $\pm$ 0.001   & 0.527 $\pm$ 0.001   & -0.520 $\pm$ 0.001  & 0.500 $\pm$ 0.002   & -0.499 $\pm$ 0.002  & 0.496 $\pm$ 0.002   & -0.497 $\pm$ 0.002  \\
$t_{\rm E}$ (days)        & 62.37 $\pm$ 0.06    & 63.75 $\pm$ 0.18    & 69.39 $\pm$ 0.32    & 65.33 $\pm$ 0.20    & 65.53 $\pm$ 0.15    & 64.77 $\pm$ 0.19    & 61.91 $\pm$ 0.42    \\
$s$                       & 1.346 $\pm$ 0.001   & 1.345 $\pm$ 0.001   & 1.341 $\pm$ 0.001   & 1.300 $\pm$ 0.002   & 1.301 $\pm$ 0.001   & 1.301 $\pm$ 0.002   & 1.296 $\pm$ 0.002   \\
$q$ ($10^{-3}$)           & 5.33 $\pm$ 0.04     & 5.07 $\pm$ 0.03     & 4.45 $\pm$ 0.04     & 6.97 $\pm$ 0.27     & 6.63 $\pm$ 0.05     & 5.92 $\pm$ 0.11     & 6.82 $\pm$ 0.19     \\
$\alpha$                  & 0.852 $\pm$ 0.001   & 0.864 $\pm$ 0.002   & -0.906 $\pm$ 0.002  & 0.861 $\pm$ 0.002   & -0.859 $\pm$ 0.001  & 0.837 $\pm$ 0.002   & -0.810 $\pm$ 0.005  \\
$\rho_\ast$ ($10^{-2}$)   & 1.103 $\pm$ 0.008   & 1.053 $\pm$ 0.007   & 0.968 $\pm$ 0.009   & 1.233 $\pm$ 0.031   & 1.194 $\pm$ 0.011   & 1.111 $\pm$ 0.014   & 1.207 $\pm$ 0.023   \\
$\pi_{{\rm E},N}$         & --                  & 0.118 $\pm$ 0.011   & -0.414 $\pm$ 0.016  & --                  & --                  & -0.143 $\pm$ 0.018  & 0.358 $\pm$ 0.042   \\
$\pi_{{\rm E},E}$         & --                  & -0.033 $\pm$ 0.007  & -0.069 $\pm$ 0.009  & --                  & --                  & 0.047 $\pm$ 0.007   & 0.008 $\pm$ 0.006   \\
$ds/dt$ (yr$^{-1}$)       & --                  & --                  & --                  & 0.765 $\pm$ 0.046   & 0.727 $\pm$ 0.017   & 0.669 $\pm$ 0.028   & 0.802 $\pm$ 0.033   \\
$d\alpha/dt$ (yr$^{-1}$)  & --                  & --                  & --                  & 1.284 $\pm$ 0.159   & -1.108 $\pm$ 0.019  & 0.497 $\pm$ 0.059   & -0.732 $\pm$ 0.085
\enddata
\tablecomments{
HJD'=HJD-2450000.
}
\end{deluxetable*}

We begin our analysis by exploring a standard set of solutions that involve modeling the event 
as a static binary lens. The Paczy\'{n}ski curve representing the evolution of the event for most 
of its duration is described by three parameters: the time of closest approach between the projected 
position of the source on the lens plane and the position of the lens photocenter\footnote{The 
"photocenter" refers to the center of the lensing magnification pattern. For a binary-lens with 
a projected separation between the lens components less than the Einstein radius of the lens, the 
photocenter corresponds to the center of mass. For a lens with a separation greater than the Einstein 
radius, there exist two photocenters each of which is located close to each lens component with an 
offset $q/[s(1+q)]$ toward the other lens component \citep{2009JKAS...42...39K}. In this case, the reference $t_0, u_0$ measurement 
is obtained from the photocenter to which the source trajectory approaches closest.}, $t_0$, the minimum 
impact parameter of the source, $u_0$, expressed in units of the angular Einstein radius of the lens 
($\theta_{\rm E}$), and the duration of time, $t_{\rm E}$ (the Einstein time-scale), required for the 
source to cross $\theta_{\rm E}$. The binary nature of the lens requires the introduction of three 
extra parameters: The mass ratio $q$ between the two components of the lens, their projected separation 
$s$, expressed in units of $\theta_{\rm E}$, and the source trajectory angle $\alpha$ with respect to 
the axis defined by the two components of the lens. A seventh parameter, $\rho_{\ast}$, representing 
the source radius normalized by the angular Einstein radius is also required to account for 
finite-source effects that are important when the source trajectory approaches or crosses a caustic \citep{2009MNRAS.399..219I}.

The magnification pattern produced by binary lenses is very sensitive to variations in $s, q$, 
which are the parameters that affect the shape and orientation of the caustics, and $\alpha$, 
the source trajectory angle. Even small changes in these parameters can produce extreme changes in 
magnification as they may result in the trajectory of the source approaching or crossing a caustic \citep{2006ApJ...642..842D, 2009ApJ...698.1826D}. 
On the other hand, changes in the other parameters cause the overall magnification pattern to vary 
smoothly.

To assess how the magnification pattern depends on the parameters, we start the modeling run by 
performing a hybrid search in parameter space whereby we explore a grid of $s, q, \alpha$ values 
and optimize $t_0, u_0, t_{\rm E}$ and $\rho_{\ast}$ at each grid point by $\chi^2$ minimization 
using Markov Chain Monte Carlo (MCMC). Our grid limits are set at $-1 \le \log s \le 1$, 
$-5\le \log q \le 1$, and $0 \le \alpha < 2\pi$, which are wide enough to guarantee that all 
local minima in parameter space have been identified. An initial MCMC run provides a map of the 
topology of the $\chi^2$ surface, which is subsequently further refined by gradually narrowing 
down the grid parameter search space \citep{2012ApJ...746..127S, 2013ApJ...763...67S}. Once we 
know the approximate locations of the local minima, we perform a $\chi^2$ optimization using 
all seven parameters at each of those locations in order to determine the refined position of 
the minimum. From this set of local minima, we identify the location of the global minimum 
and check for the possible existence of degenerate solutions. We find no other solutions.

\begin{figure}[bh]
\epsscale{1.20}
\plotone{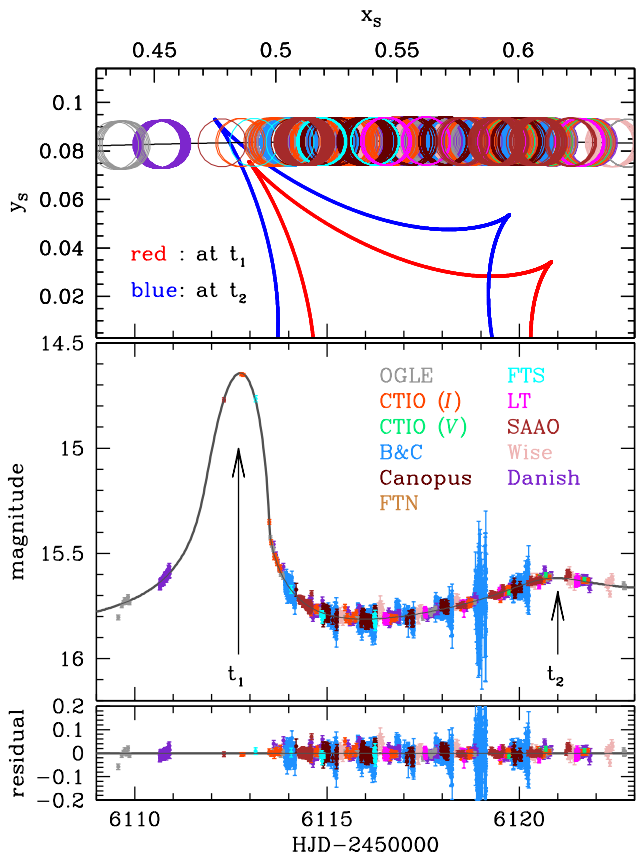}
\caption{\label{fig:two}
The bottom panel zooms-in on the anomalous region of the lightcurve presented in Figure~\ref{fig:one}. 
At the top panel we display the source trajectory, color coded for the individual contributions of each 
observatory, and caustic structure at two different times corresponding to the first and second peaks 
of the anomaly. All scales are normalized by $\theta_{\rm E}$, and the size of the circles corresponds 
to the size of the source. The first peak deviates the strongest. This is a result of the trajectory 
of the source grazing the cusp of the caustic at $t_1$ (HJD$\sim$2456112), shown in red. The second 
deviation at $t_2$ (HJD$\sim$2456121) is significantly weaker and is due to the source trajectory 
passing close to another cusp of the caustic, shown in blue. The differences in the shape of the 
caustic shown at $t_1$ and $t_2$ are due to the orbital motion of the lens-planet system.
}
\end{figure}

Since our analysis relies on data sets obtained from different telescopes and instruments which 
use different estimates for the reported photometric precision, we normalize the flux uncertainties 
of each data set by adjusting them as $e_i = f_i (\sigma_0^2 + \sigma_i^2)^{1/2}$, where $f_i$ is a 
scale factor, $\sigma_0$ are the originally reported uncertainties and $\sigma_i$ is an additive 
uncertainty term for each data set $i$. The rescaling ensures that $\chi^2$ per degree of freedom 
($\chi^2$/dof) for each data set relative to the model becomes unity. Data points with very large 
uncertainties and obvious outliers are also removed in the process.

In computing finite-source magnifications, we take into account the limb-darkening of the source 
by modeling the surface brightness as $S_\lambda(\vartheta) \propto 1 - \Gamma_\lambda 
(1 - 1.5\cos\vartheta)$ \citep{2001ApJ...549..759A}, where $\vartheta$ is the angle between the 
line of sight toward the source star and the normal to the source surface, and $\Gamma_\lambda$ 
is the limb-darkening coefficient in passband $\lambda$. We adopt $\Gamma_V=0.74$ and $\Gamma_I
=0.53$ from the \cite{2000A&A...363.1081C} tables. These values are based on our classification 
of the stellar type of the source, as subsequently described.

The residuals contained additional smooth 
structure that the static binary model did not account for. This indicated the need to consider 
additional second-order effects. The event lasted for $\gtrsim120$ days, so the positional 
change of the observer caused by the orbital motion of the Earth around the Sun may have 
affected the lensing magnification. This introduces subtle long-term perturbations in the event 
lightcurve by causing the apparent lens-source motion to deviate from a rectilinear trajectory 
\citep{1992ApJ...392..442G,1995ApJ...454L.125A}. Modeling this parallax effect requires the 
introduction of two extra parameters, $\pi_{{\rm E},N}$ and $\pi_{{\rm E},E}$, representing the 
components of the parallax vector $\pivec_{\rm E}$ projected on the sky along the north and east 
equatorial axes respectively. When parallax effects are included in the model, we use the geocentric formalism of \cite{2004ApJ...606..319G} which ensures that the parameters $t_0$, $u_0$ and $t_{\rm E}$ will be almost the same as when the event is fitted without parallax.

An additional effect that needs to be considered is the orbital motion of the lens system. The 
lens orbital motion causes the shape of the caustics to vary with time. To a first order 
approximation, the orbital effect can be modeled by introducing two extra parameters that 
represent the rate of change of the normalized separation between the two lensing components 
$ds/dt$ and the rate of change of the source trajectory angle relative to the caustics 
$d\alpha/dt$ \citep{2000ApJ...534..894A}.

We conduct further modeling considering each of the higher-order effects separately and also 
model their combined effect. Furthermore, for each run considering a higher-order effect, we 
test models with $u_0>0$ and $u_0<0$ that form a pair of degenerate solutions resulting from 
the mirror-image symmetry of the source trajectory with respect to the binary-lens axis. 
For each model, we repeat our calculations starting from different initial positions 
in parameter space to verify that the fits converge to our previous solution and that 
there are no other possible minima.

Table~\ref{table:two} lists the optimized parameters for the models we considered. We find 
that higher-order effects contribute strongly to the shape of the lightcurve. The model 
including the parallax effect provides a better fit than the standard model by $\Delta\chi^2=243.3$. 
The orbital effect also improves the fit by $\Delta\chi^2=512.8$. The combination of both 
parallax and orbital effects improves the fit by $\Delta\chi^2=563.3$. Due to the $u_0>0$ 
and $u_0<0$ degeneracy, there are two solutions for the orbital motion + parallax model which 
have similar $\chi^2$ values. Models involving the xallarap effect (source orbital motion) were also considered but they did not outperform equivalent models involving only parallax.

In Figure~\ref{fig:one}, we present the best-fit model lightcurve superposed on the observed 
data. Figure~\ref{fig:two} displays an enlarged view of the perturbation region of the 
lightcurve along with the source trajectory with respect to the caustic. The follow-up 
observations cover critical features of the perturbation regions that were not covered by 
the survey data. We note that the caustic varies with time and thus we present the shape of 
the caustic at the times of the first ($t_1$=HJD$\sim$2456112) and second perturbations 
($t_2$=HJD$\sim$2456121). The source trajectory grazes the caustic structure at $t_1$ causing 
a substantial increase in magnification. As the caustic structure and trajectory evolve with 
time, the trajectory approaches another cusp at $t_2$, but does not cross it. This second 
approach causes an increase in magnification which is appreciably lower than that of the 
first encounter at $t_1$. The source trajectory is curved due to the combination of the 
parallax and orbital effects.

The mass and distance to the lens are determined by
\begin{equation}
M_{\rm tot}={\theta_{\rm E}\over \kappa \pi_{\rm E}}; \qquad 
D_{\rm L}={{\rm AU}\over \pi_{\rm E}\theta_{\rm E}+\pi_{\rm S}},
\end{equation}
where $\kappa=4G/(c^2{\rm AU})$ and $\pi_{\rm S}$ is the parallax of the source star 
\citep{1992ApJ...392..442G}. To determine these physical quantities we require the values 
of $\pi_{\rm E}$ and $\theta_{\rm E}$. Modeling the event returns the value of $\pi_{\rm E}$, 
whereas $\theta_{\rm E}=\theta_{\ast}/\rho_{\ast}$ depends on the angular radius of the source 
star, $\theta_{\ast}$, and the normalized source radius, $\rho_{\ast}$, which is also returned 
from modeling (see Table~\ref{table:two}). Therefore, determining $\theta_{\rm E}$ requires 
an estimate of $\theta_{\ast}$.

To estimate the angular source radius, we use the standard method described in 
\cite{2004ApJ...603..139Y}. In this procedure we first measure the dereddened color and 
brightness of the source star by using the centroid of the giant clump as a reference 
because its dereddened magnitude $I_{0,c}=14.45$ \citep{2013ApJ...769...88N} and color 
$(V-I)_{0,c}=1.06$ \citep{2011A&A...533A.134B} are already known. For this calibration, 
we use a color-magnitude diagram obtained from CTIO observations in the $I$ and $V$ bands. 
We then convert the $V-I$ source color to $V-K$ using the color-color relations from 
\cite{1988PASP..100..1134} and the source radius is obtained from the $\theta_{\ast}$-$(V-K)$ 
relations of \cite{2004A&A..428..587}. We derive the dereddened magnitude and color of the 
source star as $I_0=14.62$ and $(V-I)_0=1.12$ respectively. This confirms that the source 
star is an early K-type giant. The estimated angular source radius is $\theta_{\ast}=
5.94\pm0.51$\ $\mu$as. Combining this with our evaluation of $\rho_{\ast}$, we obtain 
$\theta_{\rm E}=0.53\pm0.05$\ mas for the angular Einstein radius of the lens. 

\begin{figure}[th]
\epsscale{0.8}
\plotone{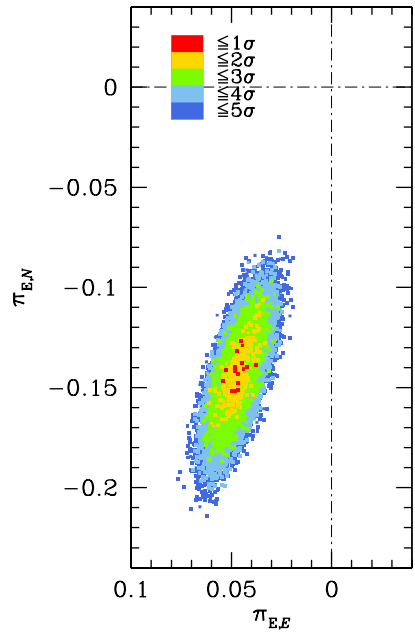}
\caption{\label{fig:three}
$\Delta\chi^2$ contours for the parallax parameters derived from our MCMC fits for the best binary-lens model 
including orbital motion and the parallax effect.
}
\end{figure}

Our analysis is consistent with the results of \cite{POLESKI}. We confirm that the lens is a planetary system composed of 
a giant planet orbiting a low-mass star and we report the refined parameters of the system. 
\cite{POLESKI} reported that there existed a pair of degenerate solutions with $u_0>0$ and $u_0<0$, 
although the positive $u_0$ solution is slightly preferred with $\Delta\chi^2=13.6$. 
We find a consistent result that the positive $u_0$ solution is preferred but the degeneracy is 
better discriminated by $\Delta\chi^2=23.7$.

\begin{figure*}[th]
\epsscale{0.9}
\plotone{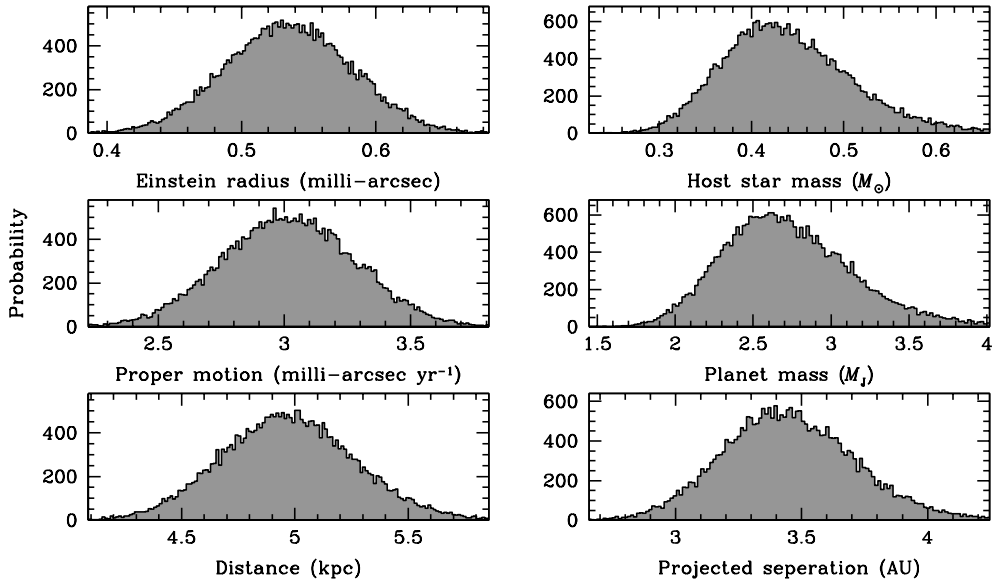}
\caption{\label{fig:four}
The physical parameter uncertainties pertaining to the lens as derived from the MCMC runs optimizing 
our binary-lens model including parallax and orbital motion for the $u_0>0$ trajectory.
}
\end{figure*}

The error contours of the parallax parameters for the best-fit model are presented in Figure~\ref{fig:three}. 
The uncertainty of each parameter is determined from the distribution of MCMC chain, 
and the reported uncertainty corresponds to the standard deviation of the distribution. 
We list the physical parameters of the system in Table~\ref{table:three} and their 
posterior probability distributions are shown in Figure~\ref{fig:four}.  

The lens lies $D_{\rm L}=4.97\pm0.29$\ kpc away in the direction of the Galactic 
Bulge. The more massive component of the lens has mass $M_{\star}=0.44\pm0.07$\ 
$M_{\odot}$ so it is an early M-type dwarf star and its companion is a super-Jupiter 
planet with a mass $M_{\rm p}=2.73\pm0.43$\ $M_{\rm J}$. The projected separation 
between the two components of the lens is $d_\perp=3.45\pm0.26$\ AU. The geocentric 
relative proper motion between the lens and the source is $\mu_{\rm Geo}=\theta_{\rm E}
/t_{\rm E}=3.02 \pm 0.26$\ milli-arcsec yr$^{-1}$. In the Heliocentric frame, the 
proper motion is $\muvec_{\rm Helio}=(\mu_{N},\mu_{E})=(-2.91\pm0.26,1.31\pm0.16)$\ milli-arcsec yr$^{-1}$.

\begin{deluxetable}{ll}
\tablecaption{Physical Parameters\label{table:three}}
\tablewidth{0pt}
\tablehead{
\multicolumn{1}{c}{parameters} &
\multicolumn{1}{c}{quantity} 
}
\startdata
Mass of the host star ($M_{\star}$)          & 0.44 $\pm$ 0.07 $M_{\odot}$         \\ 
Mass of the planet ($M_{\rm p}$)             & 2.73 $\pm$ 0.43 $M_{\rm J}$         \\ 
Distance to the lens ($D_{\rm L}$)           & 4.97 $\pm$ 0.29 kpc                 \\ 
Projected star-planet separation ($d_\perp$) & 3.45 $\pm$ 0.26 AU                  \\ 
Einstein radius ($\theta_{\rm E}$)           & 0.53 $\pm$ 0.05 milli-arcsec        \\ 
Geocentric proper motion ($\mu_{\rm Geo}$)   & 3.02 $\pm$ 0.26 milli-arcsec yr$^{-1}$
\enddata                             
\end{deluxetable}

We note that the derived physical lens parameters are somewhat different 
from those of \cite{POLESKI}.  Specifically, the mass of the host star derived in 
\cite{POLESKI} is 0.59 $M_\odot$, which is $\sim$ 34\% greater than our estimate.
Half of this difference comes from the slightly larger Einstein radius obtained by \cite{POLESKI} from the OGLE-IV 
photometry and the remaining part from the slightly larger $\pi_{{\rm E},N}$ component of the parallax obtained from  modeling the survey and follow-up photometry as presented in this paper. It should be noted that the parameters derived by both our and the\cite{POLESKI} models are consistent within the 1-$\sigma$ level.

To further check the consistency between our model and that of \cite{POLESKI}, we conducted additional modeling based on different combinations of data sets. We first test a model based on OGLE data exclusively
in order to see whether we can retrieve the physical parameters reported in \cite{POLESKI}.  
From this modeling, we derive physical parameters consistent with those of \cite{POLESKI}, 
indicating that the differences are due to the additional coverage provided by the follow-up observations. 
We conducted another modeling run using OGLE observations but also included CTIO, FTS and SAAO data, 
i.e. those datasets covering the anomalous peak. This modeling run resulted in physical parameters 
that are consistent with the values extracted from fitting all combined data together, as reported in this paper. 
This indicates that the differences between \cite{POLESKI} and this analysis, although consistent within the 1-$\sigma$ level, come mainly from follow-up data that provide better coverage of the perturbation. Therefore, using survey and follow-up data together, we arrive at a more accurate determination of the $\rho$ and $\pi_{{\rm E},N}$ parameters, which leads to a refinement of the physical parameters of the planetary system.

\section{Conclusions}
Microlensing event OGLE-2012-BLG-0406 was intensively observed by survey and follow-up groups 
using 10 different telescopes around the world. Anomalous deviations observed in the lightcurve 
were recognized to be due to the presence of a planetary companion even before the event reached 
its central peak. The anomalous behavior was first identified and assessed automatically via software agents. Most follow-up teams responded to these alerts by adjusting their observing 
strategies accordingly. This highlights the importance of circulating early models to the 
astronomical community that help to identify important targets for follow-up observations 
\citep{2012ApJ...760..116S}. There are $\sim$100 follow-up alerts circulated annually, $\sim$10\% of which turn out to be planet candidates.

Our analysis of the combined data is consistent with the results of \cite{POLESKI} and we report 
the refined parameters of the system. We find that this refinement is mainly due to follow-up observations over the anomaly.
The primary lens with mass $M_{\star}=0.44\pm0.07$\ 
$M_{\odot}$ is orbited by a planetary companion with mass $M_{\rm p}=2.73\pm0.43$\ $M_{\rm J}$ 
at a projected separation of $d_\perp=3.45\pm0.26$\ AU. The distance to the system is 
$D_{\rm L}=4.97\pm0.29$\ kpc in the direction of the Galactic Bulge.

This is the fourth cold super-Jupiter planet around a low-mass star discovered by 
microlensing \citep{2009ApJ...695..970D, 2011A&A...529A.102B, 2012ApJ...755..102Y} 
and the first such system whose characteristics were derived solely from microlensing 
data, without considering any external information.

Microlensing is currently the only way to obtain high precision mass measurements for this type of system. Radial velocity, in addition to the ${m \sin{\rm i}}$ degeneracy, at present does not have long enough data streams to measure the parameters of such systems. However, recently \cite{TRENDS-IV} have developed a promising new method to discover them using a combination of radial velocity and direct imaging. They identify long term trends in radial velocity data and use adaptive optics imaging to rule out the possibility that these are due to stars. This means that the trends are either due to large planets or brown dwarfs. This approach does not yield precise characterization but provides important statistical information. Their results are consistent with gravitational microlensing estimates of planet abundance in that region of parameter space.

The precise mechanism of how such large planets form and evolve around low mass stars is still an open question.  Radial velocity and transit surveys have been finding massive gas-giant planets around FGK-stars for years \citep{2013ApJS..204...24B}  but these stars have protoplanetary disks that are sufficiently massive to allow the formation of super-Jupiter planets. On the other hand, protoplanetary disks around M-dwarfs have masses of only a few Jupiter mass so massive gas giants should be relatively hard to produce \citep{2013AN....334...57A}. 

Recent observational studies have revealed that protoplanetary disks are as common around low mass stars as higher-mass stars \citep{2011ARA&A..49...67W}, arguing for the same formation processes. In addition, there is mounting evidence, but not yet conclusive, that disks last much longer around low-mass stars \citep{2013AN....334...57A}. Longer disk lifetimes may be conducive to the formation of super-Jupiters. The microlensing discoveries suggest that giant planets around low-mass stars may be as common as around higher-mass stars but may not undergo significant migration \citep{2010ApJ...720.1073G}.

Simulations using the core accretion formalism can produce such planets within reasonable disk lifetimes of a few Myr \citep{2012A&A...547A.112M} provided the core mass is sufficiently large or the opacity of the planet envelope during gas accretion is decreased by assuming that the dust grains have grown to larger sizes than the typical interstellar values (R. Nelson, private communication). Furthermore, gravitational instability models of planet formation can also potentially produce such objects when the opacity of the protoplanetary disk is low enough to allow local fragmentation at greater distances from the host star, and subsequently migrating the planet to distances of a few AU.

It is worth noting that highly magnified microlensing events involving extended stellar sources may produce appreciable polarization signals \citep{2012MNRAS.426.1496I}. If such signals are observed during a microlensing event,  they can be combined with photometric observations to place further constraints on the lensing geometry and physical properties of the lens.
\\
\acknowledgments

YT thanks the CBNU group for their advice and hospitality while in Korea.
DMB, MD, KH, CS, RAS, KAA, MH and YT are supported by NPRP grant NPRP-09-476-1-78 from the Qatar 
National Research Fund (a member of Qatar Foundation).
CS received funding from the European Union Seventh Framework Programme (FP7/2007-2013) under grant agreement no. 268421.
KH is supported by a Royal Society Leverhulme Trust Senior Research Fellowship.
JPB and PF acknowledge the financial support of Programme National de Planétologie and of IAP.
The OGLE project has received funding from the European Research
Council under the European Community's Seventh Framework Programme
(FP7/2007-2013) / ERC grant agreement no. 246678 to AU.
Work by CH was supported by Creative Research Initiative 
Program (2009-0081561) of National Research Foundation of Korea.
The MOA experiment was supported by grants JSPS22403003 and JSPS23340064.
TS acknowledges the support JSPS24253004. 
TS is supported by the grant JSPS23340044. 
TCH acknowledges support from KRCF via the KRCF Young Scientist Fellowship program and financial support from KASI grant number 2013-9-400-00.
YM acknowledges support from JSPS grants JSPS23540339 and JSPS19340058.
AG and BSG acknowledge support from NSF AST-1103471.
MR acknowledges support from FONDECYT postdoctoral fellowship No3120097.
BSG, AG, and RWP acknowledge support from NASA grant NNX12AB99G.
YD, AE and JS acknowledge support from the Communaute francaise de Belgique - Actions de recherche concertees - Academie Wallonie-Europe.
This work is based in part on data collected by MiNDSTEp with the Danish 1.54m telescope at the ESO La Silla Observatory. The Danish 1.54m telescope is operated based on a grant from the Danish Natural Science Foundation (FNU).

\end{document}